\documentstyle[epsf]{mn}
\def\epsfsize#1#2{\hsize}
\newif\ifAMStwofonts

\ifoldfss
  \ifCUPmtlplainloaded \else
    \NewTextAlphabet{textbfit} {cmbxti10} {}
    \NewTextAlphabet{textbfss} {cmssbx10} {}
    \NewMathAlphabet{mathbfit} {cmbxti10} {} 
    \NewMathAlphabet{mathbfss} {cmssbx10} {} 
  \fi
  \ifAMStwofonts
    \ifCUPmtlplainloaded \else
      \NewSymbolFont{upmath} {eurm10}
      \NewSymbolFont{AMSa} {msam10}
      \NewMathSymbol{\upi}     {0}{upmath}{19}
      \NewMathSymbol{\umu}     {0}{upmath}{16}
      \NewMathSymbol{\upartial}{0}{upmath}{40}
      \NewMathSymbol{\leqslant}{3}{AMSa}{36}
      \NewMathSymbol{\geqslant}{3}{AMSa}{3E}

       \let\le=\leqslant
       
    \fi
  \fi
\fi 

\ifnfssone
  \newmathalphabet{\mathit}
  \addtoversion{normal}{\mathit}{cmr}{m}{it}
  \addtoversion{bold}{\mathit}{cmr}{bx}{it}
  \newmathalphabet{\mathbfit} 
  \addtoversion{normal}{\mathbfit}{cmr}{bx}{it}
  \addtoversion{bold}{\mathbfit}{cmr}{bx}{it}
  \newmathalphabet{\mathbfss} 
  \addtoversion{normal}{\mathbfss}{cmss}{bx}{n}
  \addtoversion{bold}{\mathbfss}{cmss}{bx}{n}
  \ifAMStwofonts
    \ifCUPmtlplainloaded \else
      \UseAMStwoboldmath
      \makeatletter
      \new@mathgroup\upmath@group
      \define@mathgroup\mv@normal\upmath@group{eur}{m}{n}
      \define@mathgroup\mv@bold\upmath@group{eur}{b}{n}
      \edef\UPM{\hexnumber\upmath@group}
      \new@mathgroup\amsa@group
      \define@mathgroup\mv@normal\amsa@group{msa}{m}{n}
      \define@mathgroup\mv@bold\amsa@group{msa}{m}{n}
      \edef\AMSa{\hexnumber\amsa@group}
      \makeatother
      \mathchardef\upi="0\UPM19
      \mathchardef\umu="0\UPM16
      \mathchardef\upartial="0\UPM40
      \mathchardef\leqslant="3\AMSa36
      \mathchardef\geqslant="3\AMSa3E

       \let\le=\leqslant

    \fi
  \fi
\fi 

\ifnfsstwo
  \DeclareMathAlphabet{\mathbfit}{OT1}{cmr}{bx}{it}
  \SetMathAlphabet\mathbfit{bold}{OT1}{cmr}{bx}{it}
  \DeclareMathAlphabet{\mathbfss}{OT1}{cmss}{bx}{n}
  \SetMathAlphabet\mathbfss{bold}{OT1}{cmss}{bx}{n}
  \ifAMStwofonts
    \ifCUPmtlplainloaded \else
      \DeclareSymbolFont{UPM}{U}{eur}{m}{n}
      \SetSymbolFont{UPM}{bold}{U}{eur}{b}{n}
      \DeclareSymbolFont{AMSa}{U}{msa}{m}{n}
      \DeclareMathSymbol{\upi}{0}{UPM}{"19}
      \DeclareMathSymbol{\umu}{0}{UPM}{"16}
      \DeclareMathSymbol{\upartial}{0}{UPM}{"40}
      \DeclareMathSymbol{\leqslant}{3}{AMSa}{"36}
      \DeclareMathSymbol{\geqslant}{3}{AMSa}{"3E}

       \let\le=\leqslant

    \fi
  \fi
\fi 

\ifCUPmtlplainloaded \else
  \ifAMStwofonts \else 
    \def\upi{\pi}
    \def\umu{\mu}
    \def\upartial{\partial}
  \fi
\fi

\title{Multiple stellar population in the Sextans dwarf spheroidal 
galaxy?}
\author[M. Bellazzini, F.R. Ferraro and E. Pancino]
       {M. Bellazzini$^1$, F.R. Ferraro$^1$ 
       and E. Pancino$^2$\thanks{On leave from Dip. di Astronomia, Univ. di 
Bologna, Via Ranzani 1, I-40127, Bologna, 
	Italy}
	\thanks{Based on data taken at the European Southern
Observatory, La Silla, Chile, as part of the ESO observing programme 62.L-0354}
       \\
       $^1$Osservatorio Astronomico di Bologna, Via Ranzani 1, I-40127, Bologna, 
	Italy\\
	$^2$European Southern Observatory, Karl 
       Schwarzschild Strasse 2, D-85748 Garching bei M\"unchen, Germany}
	
\date{Submitted to MNRAS, 7 May 2001; Revised, 9 July 2001}

\pagerange{\pageref{firstpage}--\pageref{lastpage}}
\pubyear{2001}

\begin{document}

\maketitle

\label{firstpage}

\begin{abstract}
We present wide field ($33\arcmin \times 34\arcmin$) multiband (BVI)
CCD photometry (down to $I\le 20.5$) of the very low surface
brightness dwarf Spheroidal (dSph) galaxy Sextans. In the derived
Color Magnitude Diagrams we have found evidences suggesting the
presence of multiple stellar populations in this dSph. In particular
we discovered: {\it (i)} a Blue Horizontal Branch (HB) tail that appears to
lie on a brighter sequence with respect to the prominent Red HB and the
RR Lyrae stars, very similar to what found by Majewski et al. (1999)
for the Sculptor dSph; {\it (ii)} hints of a bimodal distribution in color of
the RGB stars; {\it (iii)} a double RGB-bump.  All
these features suggest that (at least) two components are present in
the old stellar population of this galaxy: a main one with
$[Fe/H]\sim -1.8$ and a minor component around $[Fe/H]\sol -2.3$.
The similarity with the Sculptor case may indicate that multiple star formation
episodes are common also in the most nearby dSphs that ceased their
star formation activity at very early epochs.
\end{abstract}


\section{Introduction}

The Sextans dwarf spheroidal (dSph) galaxy is the least studied member
of the family of dwarf galaxies that orbits the Milky Way. Discovered
just a decade ago by Irwin et al. (1990), it is the Local Group galaxy
with the {\em lowest} surface brightness ($\mu_0=26.2\pm 0.5$
mag/arcsec$^2$), it is relatively distant ($D=86\pm 4$ Kpc; data from
Mateo 1998), and it is projected onto a sky field significantly
contaminated by foreground stars belonging to our Galaxy ($b=+42.27$).
The body of the Sextans dSph is devoid of gas, but an HI cloud
($M \simeq 3\times 10^4 ~M_{\sun})$ with similar velocity has
been recently discovered by Blitz \& Robishaw (2000), $\sim 2 \degr$
apart from the center of the galaxy.

Mateo et al. (1991) presented the first Color Magnitude Diagram (CMD)
of Sextans based on CCD photometry. They obtained deep photometry of
two overlapping $\sim 3.9\arcmin \times 3.9\arcmin$ fields ($V<24$,
barely reaching the Turn Off point), and a shallow photometry of nine
similar fields.  They derived a ``globular cluster like'' CMD, with a
predominantly red Horizontal Branch (HB) and a large number of Blue
Straggler Stars (BSS). From the analysis of the CMD, they
concluded that the galaxy is dominated by a very old ($\sim 15$ Gyr)
population with a metallicity of $[Fe/H]\sim -1.6 \pm 0.2$.  Da Costa
et al. (1991) obtained spectra of 14 stars in Sextans, from which they
derived the first estimate of the systemic radial velocity,
identifying 6 member stars, and a mean metal abundance of
$[Fe/H]=-1.7\pm 0.25$ from the CaII triplet, in agreement with Mateo
et al. (1991).  Both teams noted that this metallicity was
significantly higher than expected given the observed integrated
magnitude ($M_V\simeq -9.5$) and the relation between [Fe/H] and $M_V$
followed by the other dSphs (Mateo et al. 1998).

\begin{figure*}
\vspace{20pt}
\epsffile{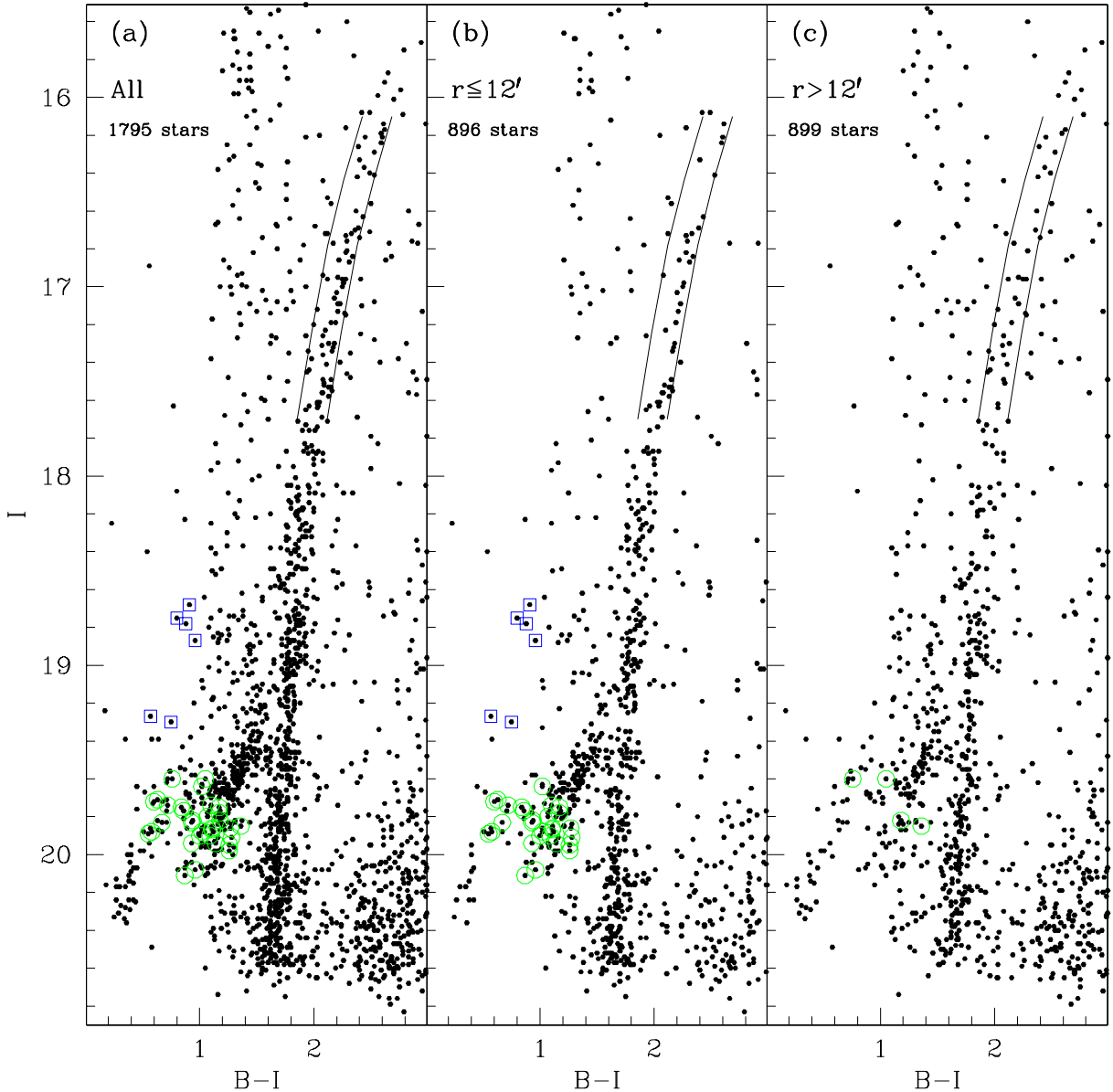}
\caption{(I, B-I) Color Magnitude Diagrams for the Sextans dSph.
Variables (from MFK) are marked with different symbols: RR Lyrae ({\it
large open circles}) and anomalous Cepheids ({\it large open
squares}). The typical photometric errors ($\epsilon_I,\epsilon_{B-I}$)
range from (0.01,0.015) at $I=16$ to (0.03,0.04) at $I=20.5$. Panel {\em (a)}:
CMD of the entire sample; panel {\em (b)}: stars within $12\arcmin$ from the
center of the galaxy; panel {\em (c)}: stars with $r > 12\arcmin$.
The two parallel lines approximately enclose
the color range spanned by Sex RGB stars to show more clearly the different
color distribution of inner and outer RGBs.
The position of the center has been assumed to coincide with the star V7 by MFK,
accordingly to their prescriptions. }
\end{figure*}

Suntzeff et al. (1993; hereafter S93) obtained spectra for 80 stars in
the direction of the Sextans dSph, identifying 43 member stars for which
they found $-2.44\lid [Fe/H] \lid -1.57$ from the CaII triplet (see
Da Costa et al. 1991).  S93 concluded that {\it (a)} the mean
metallicity of Sextans is $[Fe/H]=-2.05\pm 0.04$, fully compatible
with the $M_V$ vs. $[Fe/H]$ relation of dSphs, and that {\it (b)}
there is an intrinsic metallicity dispersion of $0.19$ dex.  Mateo,
Fischer \& Krzemi\'{n}ski (1995; hereafter MFK) performed a variable
stars survey over a large area ($18\arcmin \times 18\arcmin$),
identifying 36 RR Lyrae and 6 anomalous Cepheids, all probable members
of the Sextans galaxy. They confirmed that the dominant stellar
population in Sextans is old ($> 10$ Gyr) and they associated the
previously detected BSS with a younger ($2-4$ Gyr) population,
comprising less than $25 \%$ of the whole stellar content.  MFK
estimated the overall Sextans metallicity to be $[Fe/H]\simeq -1.6$,
from both the Red Giant Branch (RGB) color and the RR Lyrae stars
properties.  They also measured the HB morphology parameter
($P_{HB}=(B-R)/(B+R+V)$),\footnote{$B$, $R$ and $V$ are respectively:
the number of stars bluer than the instability strip ($B$), redder
than the instability strip ($R$) and lying inside the instability
strip, i.e., the RR Lyrae variables ($V$).} obtaining $P_{HB}=-0.5$
and they noted that this value is not compatible with an old age and a
metallicity as low as that found by S93. On the other hand, if
$[Fe/H]\simeq -1.6$ is assumed, the HB morphology parameter turns out
to be consistent with an age of $\sim 12$ Gyr.

Geisler \& Sarajedini (1996) derived $[Fe/H]=-2.0\pm 0.1$, with an
intrinsical spread of $0.17$ dex, from Washington photometry of RGB
stars, in full agreement with S93. Finally, Shetrone, C\^{o}t\'{e} \&
Sargent (2001; hereafter SCS) obtained Keck-HIRES spectra of five
giants in Sextans, for which they derived $[Fe/H]=-2.85$, $-2.19$,
$-1.93$, $-1.93$ and $-1.45$. 

In conclusion, there is no agreement on the actual metallicity of this
dSph, and indeed serious difficulties emerged in the interpretation of
its HB morphology, at least if a classical ``single age - single
metallicity'' scheme is adopted for the dominant population (see MKF).

The advent of modern wide field CCD cameras has significantly changed
our view of the dSph galaxies, once believed to be prototypes of Pop
II systems. Deeper photometry and larger samples have revealed that
most of these systems had complex and various star formation and/or
chemical enrichment histories (see Mateo 1998, van den Bergh 1999,
Grebel 1999, and references therein).  This can be the case also for
the Sextans dSph. Given its extremely low surface brightness, it is
quite possible that relevant features of the CMD have been missed,
since even when observing relatively wide fields, some short-lived
evolutionary sequences may not be sufficiently sampled.  For example,
the largest published photometric sample (MKF) for Sextans includes
only 505 stars with $V\le 21$ (i.e., $\sim 0.5$~mag below the HB), a
significant fraction of them being foreground or background sources.

In this Letter we present wide field ($\sim 34\arcmin \times
33\arcmin$) BVI photometry (down to $V\sim 21.5$) in the Sextans
galaxy and we discuss the main features of the CMD. The area covered
by our study is $\sim 3.5$ times larger than the largest survey (MKF)
published for this galaxy, so far. From the analysis of the main 
branches in the CMD we find several evidences suggesting the presence of two
components in the main stellar population of the Sextans dSph.
\section{The Color Magnitude Diagram}

One B and one I images (both $600~s$-long) of the Sextans dSph 
have been obtained in January 1999 at the 2.2m
ESO-MPI telescope at La Silla (Chile), using the Wide Field Imager
(WFI), a mosaic camera with a total field of view of $\sim 34\arcmin
\times 33\arcmin$. The observations were carried on as a back-up programme
while the main targets were not visible. The seeing was $\sim 1.2 \arcsec$ FWHM.
The observational set-up and the
photometric calibration are the same as in Pancino et al. (2000;
hereafter P00).
 
The standard bias and flat-field correction has been applied to the
data, using the NOAO {\em mscred} package in IRAF\footnote{IRAF is
distributed by the National Optical Astronomy Observatories, which is
operated by the association of Universities for Research in Astronomy,
Inc., under contract with the National Science Foundation.}. PSF
photometry was then performed using DoPHOT (Schechter, Mateo \& Saha 1993).
 
The observed field is remarkably contaminated by background
galaxies. We retained only the sources classified as bona fide stars
by DoPHOT. The automated star/galaxy classification by DoPHOT was
independently (and successfully) checked using the SExtractor package
(Bertin \& Arnouts 1996).  The test showed that our sample is not
significantly contaminated by background galaxies for $I\sol 19$,
while no useful check was possible at fainter magnitudes.

The panel {\em (a)} of Fig.~1 shows the $(I,B-I)$ CMD for $\sim 1800$ sources 
detected in the WFI field of view.  
The steep RGB and the extended HB of the Sextans
galaxy are clearly stacking over the nearly uniform background in the
CMD, thus showing that the lacking of a control field is not a serious problem
in the present case. We have cross-identified most of the variable stars 
observed by MFK (see Fig.~1).  
There are a number of features worth of noticing in Fig.~1a:

\begin{enumerate}

\item The HB is very extended: indeed, the most notable feature in the
CMD is the presence of a well populated blue tail (hereafter BHB) at
$I\sim 20$ and $B-I<0.8$. The BHB appears to lie above the mean
level defined by the red HB and RR Lyrae stars. The effect is of course
even more evident in the $(B, B-I)$ CMD (not shown) and strongly
resembles the anomalous HB morphology noted by Majewski et al. (1999;
hereafter M99) in the Sculptor dSph. This is the first time that this
feature is detected in Sextans.
      
\item  The RGB appears steep and well defined. However, the spread in
color of the giants is much larger than the observational error, hence
a sizeable intrinsic metallicity dispersion could be present. The RGB
thickness is clearly visible in the region $18<I<19$.

\item  The location of the bulk of the RGB stars in the $(I,B-I)$ CMD
turns out to be bluer than the location of the dominant population of
giants in $\omega$ Cen (see P00), suggesting that most of the stars in
the Sextans dSph have $[Fe/H]\sol -1.6$ (see also below).  We also
cross-identified all the member giants with known metallicity (from
S93): there is no apparent color segregation as a function of the S93
metallicity estimates. This could be due to uncertainties in
the CaII triplet measurements, and in this case high-resolution
spectroscopy could help in better understanding the actual metallicity
distribution.
 
\item As expected, the contamination by foreground and background
sources is quite large. 
The sparse blob of points at $(B-I)>2.6$ and $I>20.5$ is
probably due to the increasing number of faint and unresolved
galaxies. It is interesting to note the sharp cut off in color the 
distribution of 
foreground stars occurring at $B-I\sim 1.0$ ($B-V\sim 0.4$), which 
 corresponds to the TO color of the halo and/or thick disc of the Milky
 Way (Unavane, Wyse \& Gilmore 1996; Morrison et al. 2000). 
However, on
the basis of the position in the CMD, we estimate that $\sim 1000$
stars are likely members of the Sextans dSph: this is the largest
sample of evolved stars ever published for the Sextans dSph so far. 

\end{enumerate}

The panels {\em(b)} and {\em(c)} of Fig.~1 shows the CMD for the stars within
$12 \arcmin$ from the center of Sextans and outside this circle, respectively.
It is interesting to note that the relative abundance of BHB stars is larger in
the outer region. The ratio between the number of HB stars with $B-I\le 0.5$
($N_{BHB}$) and those with $B-I> 1.1$ ($N_{RHB}$) is  
$N_{BHB}\over{N_{RHB}}$ $=$ $13\over{109}$ $= 0.12 \pm 0.03$ for the inner 
sample and $N_{BHB}\over{N_{RHB}}$ $=$ $18\over{74}$ $= 0.24 \pm 0.06$ for the 
outer one.
Even more remarkably, the upper RGB stars with $r\le 12 \arcmin$ are tightly
gathered along the red edge of the color distribution while the outer ones are
evenly spreaded over the whole color range spanned by the total sample of RGB
(delimited in Fig.~1 by two parallel curves enclosing the RGB, reported as an 
aid for the eye). This kind of
population gradient has been observed in many dSph galaxies (Mateo 1998
and references therein; see also M99 and Hurley-Keller, Mateo \& Grebel 1999 
for the case of Scl that is particularly relevant here) and provides a first 
indication on the possible composite nature of the dominant population in 
Sextans. Unfortunately, the low total number of Sextans stars prevents further
detailed study of the inner and outer RGB samples separately.
Thus, in the following discussion, we will refer to the total RGB sample.

\section{A composite population?}

As quoted above, the most notable {\it new} feature discovered in this
Letter is the presence of an {\em anomalous} blue HB tail. 
We find that the BHB
population, defined by considering all the non-variable stars with
$(B-I)<0.8$, comprises $\sim 17\%$ of the entire HB population. With
this definition, the $P_{HB}$ parameter turns out to be
$P_{HB}\simeq-0.4$, suggesting a bluer morphology with respect to the
one estimated by MKF. Note that this value should be considered as a
lower limit, since the foreground contamination can significantly
enhance the counts in the R bin, leaving untouched the B one.
     
Some of the features noted in the CMD of Sextans have been recently
found in the CMD of another dSph: the Sculptor galaxy by M99. In
particular, they interpreted the apparent mismatch between the HB tail 
and the RR Lyrae level in Sculptor as the
signature of the presence of two populations, a very metal poor one
(at $[Fe/H]\sim -2.2$) responsible for the brighter BHB, and a more
metal rich one (at $[Fe/H]\sim -1.5$) associated with the red HB (see
Hurley-Keller et al. 1999 for a different view).  Their
hypothesis was supported also by the Luminosity Function (LF) of the
RGB, in which two distinct RGB bumps were detected.  Intrigued by the
similarity, we proceeded in a deeper analysis following
M99. Unfortunately, the extremely low surface brightness of Sextans
and the foreground contamination make our analysis more uncertain,
despite the fact that we observed a significantly wider field than
M99.

\begin{figure}
\vspace{20pt}
\epsffile{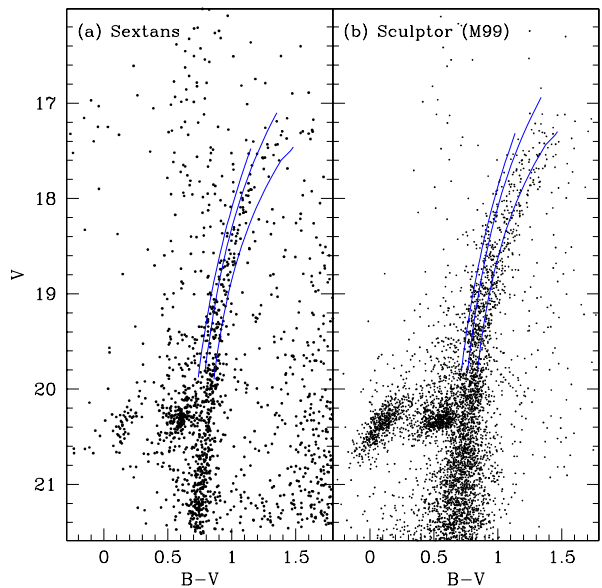}
\caption{(V, B-V) CMDs for Sextans ({\em panel (a)}, this work) and
Sculptor ({\em panel (b)}, from M99). RGB ridge lines of template
globular clusters (from Ferraro et al. (1999) are over-plotted.  From
blue to red: NGC~5053 ($[Fe/H]_{ZW}=-2.58$), M~68
($[Fe/H]_{ZW}=-2.09$), and M~3 ($[Fe/H]_{ZW}=-1.66$). The CMD presented in 
{\em panel (a)} has been obtained by coupling the photometry from our 600 $s$
B exposure with the one from 300 $s$ V exposure we retrieved from the WFI 
archive.}
\end{figure}

The CMD of Sculptor presented by M99 is in the
$(V,B-V)$ plane, for this reason, in
 order to perform a direct comparison among the various features in
the CMDs of the two dSph, we retrieved a $300~s$ $V$
exposure of Sextans (obtained in January 2000) from the WFI archive.
This image covers
approximately the same field of our $B,I$ dataset. The absolute
calibration of the $V$ data was obtained using $\sim 350$ stars in common with
MKF.

In Fig.~2 we compare our $(V, B-V)$ CMD of Sextans with the one of
Sculptor (from M99\footnote{The M99 dataset was kindly provided in
electronic form by M.H. Siegel.}). The similarity of the two CMDs, and
in particular of their {\em HB morphology} is indeed striking,
suggesting that the presence of populations with different mean
metallicities may be also the origin of the Sextans {\it anomalous} HB
morphology. 

The RGB ridge lines of three template globular clusters taken from
Ferraro et al. (1999; hereafter F99) have also been overplotted on the
CMDs, once corrected for distance modulus and reddening according to
Mateo (1998). The adopted templates are (from the bluest to the
reddest ridgeline) NGC~5053, NGC~4590 (M~68) and NGC~5272 (M~3), with
metallicities of $[Fe/H]=-2.58, -2.09$ and $-1.66$ in the Zinn \& West
(1984; hereafter ZW) scale, and of $[Fe/H]=-2.51, -1.99, -1.34$ in the
Carretta \& Gratton (1997; hereafter CG) scale, respectively. Here we
quoted the CG scale just for completeness: since most previous
estimates are in the ZW scale, we adopt the ZW metallicity scale
throughout the rest of this Letter.
 
First, we note that the vast majority\footnote{A few stars seem to run
parallel to the main RGB sequence, at the red side of the M~3
ridgeline. The membership of these stars is to be assessed
spectroscopically: they are likely to be foreground stars, but they
may also be part of a minor metal rich component of the galaxy.} of
the Sextans RGB stars lie to the blue side of the M~3 ridgeline.  This
means that $[Fe/H]_{ZW}\sim -1.6$ is, at most, an upper limit to the
metallicity distribution of the main population of the Sextans dSph.
The main fraction of the Sextans giants is enclosed between the
ridgelines of M~3 and M~68 (suggesting the presence of a main
population with $[Fe/H]\sim -1.8$), while a minor fraction ($\sim
25\%$) lies between the ridgelines of M~68 and NGC~5053 (suggesting
the presence of a lower metallicity component with $[Fe/H]< -2.1$).
These mean metallicities are in good agreement with the previous
estimates by S93, SCS and Sarajedini \& Geisler (1996), confirming the
presence of a quite large metallicity spread in Sextans ($-2.6\sol
[Fe/H]_{ZW}\sol -1.6$).  Moreover, it is interesting to note that the
most metal-poor component fraction ($\sim 25\%$) turns out to be
consistent with the BHB population fraction ($\sim 17\%$, see
above). This result further supports a possible
connection between the metal-poor RGB component and the BHB.  

\begin{figure}
\vspace{20pt}
\epsffile{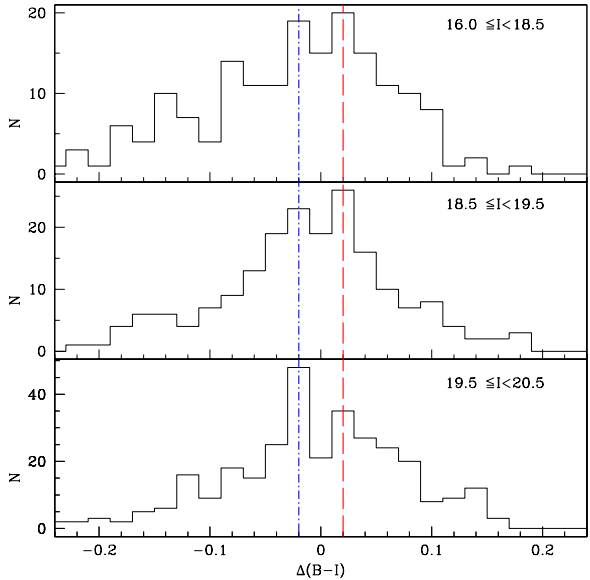}
\caption{Distribution of the color deviation from the ridge line of the 
Sextans  giants in three different ranges sampling more than $4$ mag along the
RGB. The modes of the two components are marked by vertical lines 
crossing the panels.}
\end{figure}

In order to obtain better constraints on the metallicity distribution in the
Sextans dSph we analyzed in more detail the RGB colour distribution 
shown in Fig.~1. We derived a mean ridge line for the whole RGB 
following the method adopted by F99 and we computed the color deviations
$\Delta (B-I)$ of RGB stars from the ridge line. The distribution of 
$\Delta (B-I)$ in three magnitude ranges sampling the entire
extension of the observed RGB is shown in Fig.~3. 
While the distribution is quite broad (also
because of the foreground contamination) the core of the distribution presents
two separate peaks at the same position in all the considered ranges. 
The suggested bimodal color distribution of the giants in Sextans, already
partially apparent in Fig.~1b,c, provides further support for
the presence of (at least) two stellar populations of different metallicity.

\begin{figure}
\vspace{20pt}
\epsffile{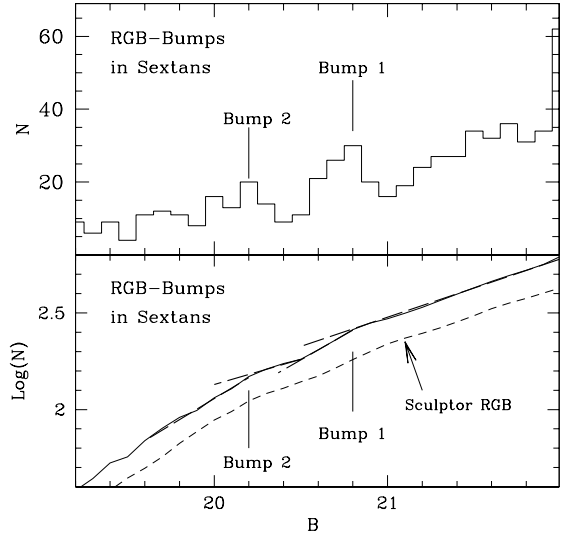}
\caption{{\em Panel (a)}: differential LF of the Sextans RGB. {\em
Panel (b)}: cumulative logarithmic LF of the same sample. The two
distinct bumps are clearly indicated. In {\em panel (b)} the LF of
Sculptor from M99 is also reported for comparison, after the
application of an arbitrary shift.}
\end{figure}

Finally, a further remarkable similarity with the case of Scl is shown in
Fig.~4, where the RGB-LFs are plotted, following the standard approach
to reveal the RGB bump position (see Fusi Pecci et al. 1990, F99 and
references therein). {\em Two distinct RGB bumps are detected along
the Sextans RGB}, the fainter one (B1) is located at $B=20.8 \pm 0.1$
(corresponding to $V\simeq 19.95$) and the brighter one at $B=20.2 \pm
0.1$ (corresponding to $V\simeq 19.35$). The luminosity of the RGB
bump depends primarily on the metal content of the associated
population and only weakly on age: more metal deficient populations
have brighter bumps. Hence the double RGB bump can be interpreted as
the signature of the coexistence of two populations of different mean
metal content, as in M99.  Assuming $(m-M)_V=19.79\pm 0.15$ from MFK,
we derive $M_V^{B1}=0.16$ and $M_V^{B2}=-0.44$, respectively.  We can
use the absolute magnitude of the bump in order to get a rough
estimate of the metallicity for the two populations, by using the
following relation: $$M_V^{bump}=0.87 [Fe/H]_{ZW} +1.74$$ which has
been derived from the F99 data-set. From the absolute magnitudes
obtained above, we finally find $[Fe/H]_{ZW}^{B1}\sim -1.8$ and
$[Fe/H]_{ZW}^{B2}\sim-2.5$\footnote{Note that this value is quite
uncertain, since the relation derived above is valid only in the range
$-0.2\sol[Fe/H]_{ZW}\sol-2.2$ and significant deviation from
linearity cannot be excluded outside this range.}

\section{Conclusion}

From all the above results, a remarkably self-consistent scenario
emerges, i.e., {\em the dominant population in Sextans consists of two
distinct components, both quite metal deficient: a minor one ($\sim
20\%$ of the whole) with $[Fe/H]_{ZW}\sol -2.3$, and the most conspicuous
one with $[Fe/H]_{ZW}\sim -1.8$}. The most metal poor population is
likely to be associated with the BHB and with the bright RGB bump B2,
while the major component should be related with the RHB and the B1
bump. As in many other dSphs the component associated with the BHB stars
shows a more extended radial distribution with respect to the RHB and the
reddest RGB stars.
This scenario is consistent with the results by S93, SCS and
Sarajedini and Geisler (1996), leaving Sextans in the expected place
in the $M_V$ vs. $[Fe/H]$ relation of dSphs.  Since it is reasonable
to presume that the metallicity difference of the two components may
be associated to an age difference (up to a few Gyr), the peculiar HB
morphology can be much more easily understood in the above framework
than in an usual ``single-metallicity'' scheme, that has been proved
to be inadequate for all the other dSphs (Mateo 1998; Grebel 1999).
However, it is important to remark that a confirmation of the above framework
has to wait for deeper photometry, reaching the Main Sequence to well below the
Turn Off point, and possibly high resolution spectroscopy of larger samples.
Lacking these fundamental pieces of information, also other interpretations may
be viable (see, f.i, M99 and Hurley-Keller et al. 1999). 

The most nearby dSph galaxies host only modest (if any)
intermediate-age populations (Mateo 1998; van den Berg 1999) and they
are almost completely gas depleted in their main
body\footnote{Sculptor, Ursa Minor, Draco and Sextans seem indeed
quite similar in stellar and gas content as well as in distance from
the center of the Galaxy and other characteristics. In particular, SCS
claim that the UMi, Dra and Sextans dSphs {\em (a)} share {\em very
similar} abundance patterns, distinct from those typical of the
Galactic halo stars, {\em (b)} their dominant stellar populations are
quite metal deficient, with no star more metal rich than
$[Fe/H]_{ZW}\sim -1.4$, and {\em (c)} many of their stars reach a
degree of metal deficiency ($[Fe/H]_{ZW}\le -2.5$) not spanned by
existing Galactic globulars.}  (Blitz \& Robishaw 2000). It has been
convincingly suggested that the interaction with the tidal field
and/or with the hot gaseous halo and/or with the early stellar wind
from the Milky Way were particularly efficient in stopping the main
evolutionary path of these innermost and least massive satellites at
early epochs (see van den Bergh 1994; Blitz \& Robishaw 2000, and
Bellazzini, Fusi Pecci \& Ferraro 1996). Despite this, at least some
of them (Sextans, Sculptor) seem to host multiple populations,
indicating that their stellar content {\em was not originated in a
single burst} and that a significant chemical evolution took place
before their gas reservoirs were swept out (see Tamura, Hirashita \& Takeuchi
2001).  
Hence, these galaxies may
provide an unique view of a very early phase of the evolution of
stellar systems, for which there is no other counterpart, except for
the rare very metal poor stars that are dispersed in the vast Galactic
halo (Cayrel 1996; Beers 1999; Morrison et al. 2000).

\section*{Acknowledgments}

This research has been funded by the Italian MURST through the COFIN
p.  {\small MM02241491\_004} grant, assigned to the project {\em
Stellar Observables of Cosmological Relevance}. We are grateful to Bob
Rood and M.H. Siegel for providing the M99 data in electronic form and
to L. Pasquini for useful discussions. We are indebted to the referee for the
constructive criticisms and the useful suggestions that enhanced the overall
quality of this Letter. E.P. aknowledges the
hospitality of the ESO Studentship Programme. Part of the data
analysis has been performed using software developed by P. Montegriffo
at the Osservatorio Astronomico di Bologna.  This research has made
use of NASA's Astrophysics Data System Abstract Service.

\label{lastpage}

\end{document}